\RequirePackage[T1]{fontenc}
\documentclass[10pt,conference]{IEEEtran}
\usepackage{cite}
\usepackage{url}
\usepackage{graphicx}
\usepackage{balance}
\usepackage{microtype}
\usepackage{xurl}
\newsavebox{\figcbox}
\usepackage[hidelinks]{hyperref}
\newcommand{\OS}{\mbox{Omo Space}}
\newcommand{\OT}{\mbox{Omo Theia}}
\title{Omo: Giving Continuing AI-Agent Work\\a Place to Return To}
\author{\IEEEauthorblockN{Kaviru Hapuarachchi}
\IEEEauthorblockA{Omo Research\\Colombo, Sri Lanka\\kaviru@omo.sh}
\and
\IEEEauthorblockN{Harrison Edwards}
\IEEEauthorblockA{Omo Research\\Melbourne, Australia\\harry@omo.sh}}
\begin{document}
\bstctlcite{IEEEexample:BSTcontrol}
\maketitle
\begin{abstract}
AI coding agents now keep working while the people who started them are away. Coming back is the hard part: a session list tells you which agent is where, but not what it was asked to do, what it changed, or what to tell it next. We present Omo, which gives each piece of continuing agent work a persistent place to return to. We built the idea twice. Omo Space is an open-source Minecraft mod in which rooms and screens that people build are live terminals on their own machine. Omo Theia brings the same rooms and desks into a customizable IDE, one step away from the editor and the Git patch. Both keep agent sessions running on the host after the view is closed. We describe the two systems, four lessons from building them, and a study designed to test whether returning to a place helps people pick up agent work correctly.
\end{abstract}
\begin{IEEEkeywords}
spatial IDEs, AI coding agents, task resumption
\end{IEEEkeywords}

\section{Introduction}
Consider a builder who asks an agent to make a worksheet generator keep line breaks in both the preview and the exported PDF, and then leaves for a meeting. An hour later the agent reports that it is done. To continue, the builder has to remember what was asked, see that only the preview was fixed, and decide on the next instruction. Finding the agent's terminal is the easy part. Reconstructing what the work was about is not.

Coming back to interrupted work has long been known to be hard. In 10,000 recorded programming sessions, programmers rarely resumed editing within a minute and usually navigated and searched for task context first~\cite{parnin}. Memory for an unfinished goal fades unless something in the environment cues it~\cite{altmann}, and people who are interrupted keep up by working faster at the cost of more stress, frustration, and effort~\cite{mark}. With agents, the interruption is built in: the work continues, and changes, while the person is away. This is becoming common, as developers run several agents at once and step away while they work~\cite{parallel}.

Existing tools handle parts of the problem. Terminal multiplexers such as tmux keep a process alive. Session lists and dashboards show which agent is running and summarize its state. None of them helps a person rebuild the intent behind a piece of work they last saw an hour ago.

To be precise about the problem, we separate three things that session persistence tends to collapse into one. We call keeping an agent's process running \emph{execution continuity}. We call finding the relevant work again \emph{workspace recognition}. We call reconstructing the goal, the relevant change, and a sensible next action \emph{context recovery}. Multiplexers solve the first. Omo is about the second and, we hypothesize, the third: if a piece of work always lives in the same place, returning to that place may bring back what you meant to do there, much as people recall material better in the environment where they learned it~\cite{godden}.

The people facing this are also changing. Most programs have long been written by people who are not professional developers~\cite{ko}, and code-generating models now let non-experts build software by describing what they want, even though learning how to direct the model is itself a challenge~\cite{liu}. Yet IDEs still assume a trained developer, who knows where to look in a terminal, a file tree, and a diff. A teacher building a worksheet tool, or a founder running agents between meetings, has none of that habit to fall back on.

Omo began with our own work. As a two-person team running several products, we found that starting agent work had become easier than keeping track of it: which agent needed an answer, where each project stood, what had already been finished. One of us (H.E.) moved from voice and hologram prototypes to Minecraft, and giving each project a room was the change that stuck. Since October 2026 we have run our content pipeline in Omo Space, one room per stage, and we meet and coordinate inside that world. The other (K.H.) built the Theia version, so that people who do not play Minecraft, or who want a real editor, can use the same idea. This is our experience, not evidence that places help anyone else, and it is why we want to test the idea.

\begin{figure*}[!t]
\centering
\begin{minipage}[t]{2.04in}\centering\includegraphics[height=1.36in]{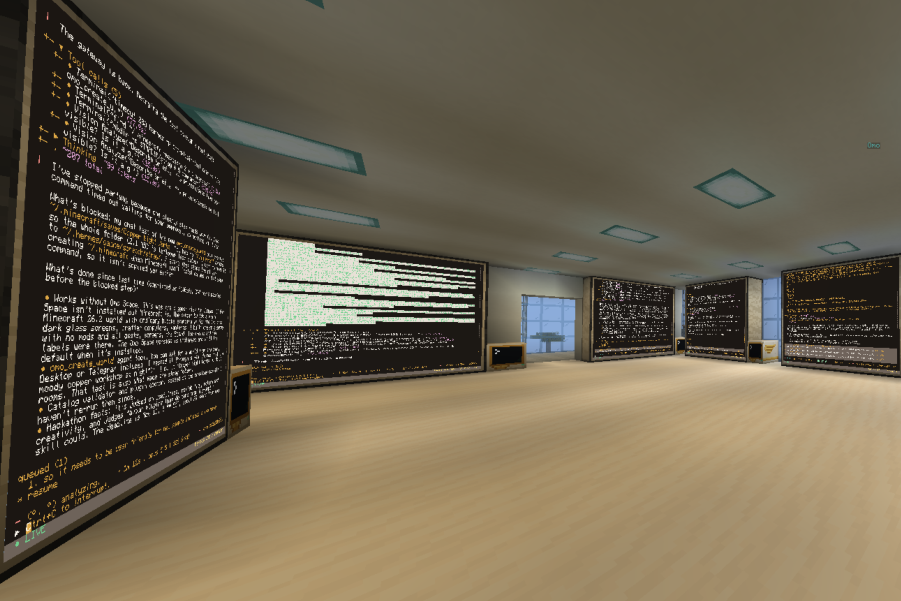}\\[1pt]{\footnotesize \textbf{(a)} \OS: a hall of agent screens}\end{minipage}\hfill\begin{minipage}[t]{4.94in}\centering\includegraphics[height=1.36in]{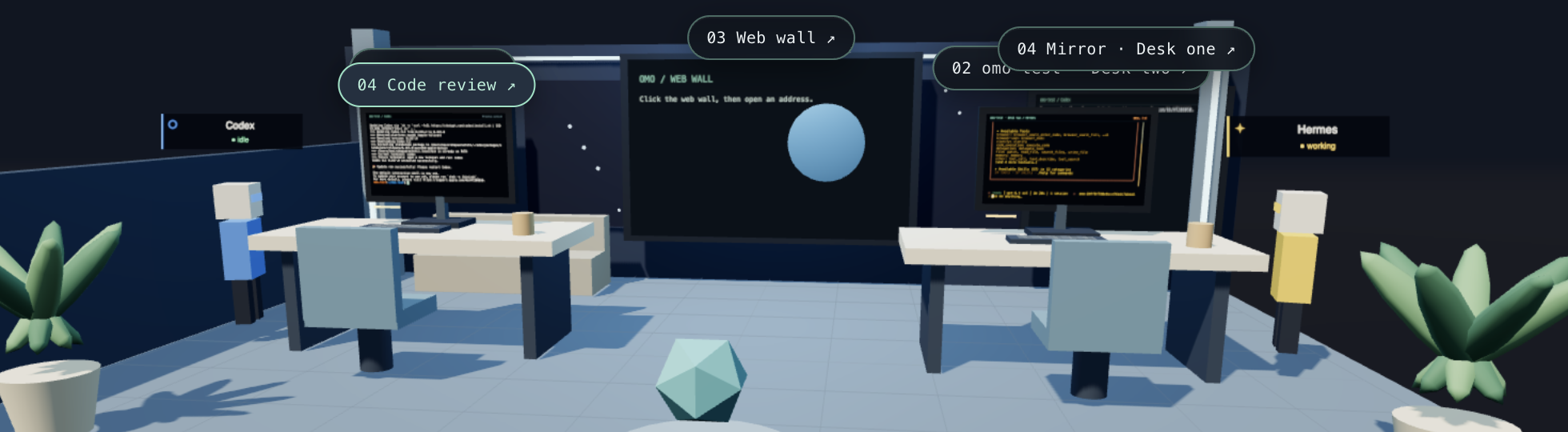}\\[1pt]{\footnotesize \textbf{(b)} \OT: one project room; Codex at Desk one, Hermes at Desk two, a web wall}\end{minipage}\\[5pt]
\begin{minipage}[t]{2.96in}\centering\includegraphics[height=1.37in]{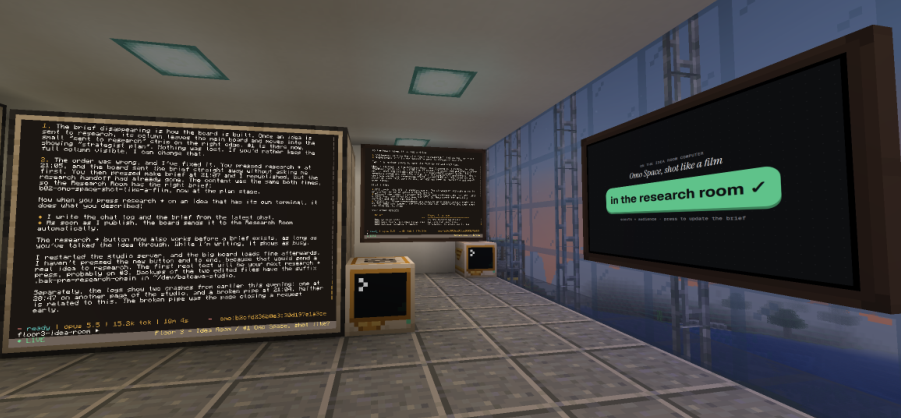}\\[1pt]{\footnotesize \textbf{(c)} \OS: an agent terminal in the Idea Room}\end{minipage}\hfill\begin{minipage}[t]{1.82in}\centering\includegraphics[height=1.37in]{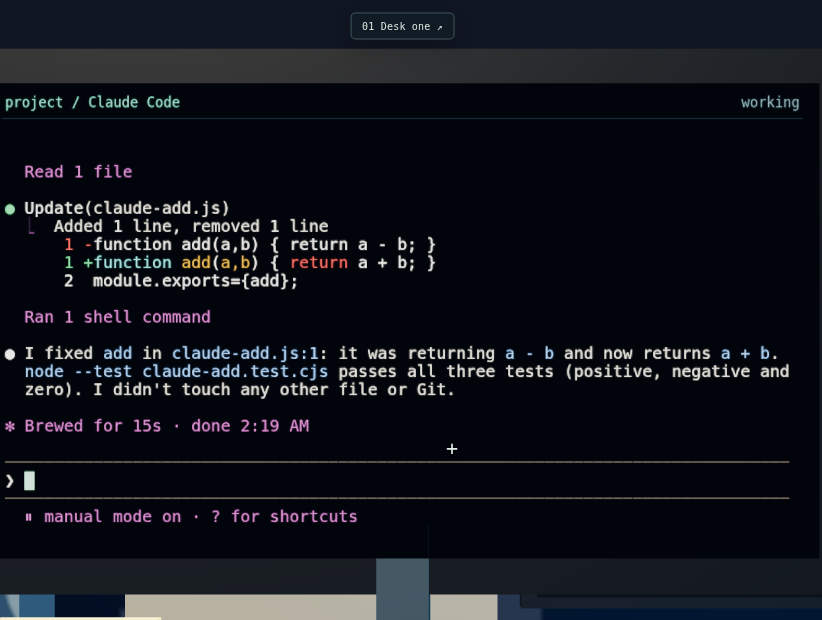}\\[1pt]{\footnotesize \textbf{(d)} \OT: Claude Code at a desk}\end{minipage}\hfill\begin{minipage}[t]{2.13in}\centering\includegraphics[height=1.37in]{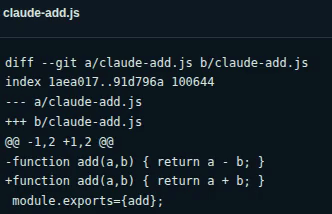}\\[1pt]{\footnotesize \textbf{(e)} \OT: Git review of the repair}\end{minipage}
\caption{One idea, two places. (a, c) \OS{}: a hall whose walls hold live agent terminals, and the Idea Room of our content pipeline, where an agent terminal sits beside a card showing that the idea has moved on to the research room. (b, d, e) \OT{}: a project room with Codex and Hermes at separate desks, Claude Code at a desk, and the Git review of the agent's change. The label in (d) still reads ``working'' while the terminal reports ``done.'' All panels are cropped screen captures. \textbf{The place shows where the work lives; only the terminal and the patch show what happened.}}
\label{fig:places}
\end{figure*}

This paper makes the following contributions:
\begin{itemize}
\item We frame returning to agent work as three separate problems and identify context recovery as the one current tools leave open.
\item We present two implementations of one idea that share a backend: \OS{}, a released Minecraft mod, and \OT{}, a spatial IDE built on Eclipse Theia (Fig.~\ref{fig:places}).
\item We report four lessons for IDE builders from building the idea twice.
\item We propose a study that tests whether places help people resume agent work correctly, against well-named sessions and a stable 2D layout.
\end{itemize}

\section{Related Work}
Using space to organize work is an old idea. Data Mountain let people arrange documents on a virtual surface and find them again by location~\cite{mountain}. Code Canvas placed a whole project on one zoomable surface and warned that changes made by others threaten spatial stability~\cite{canvas}, which is exactly what an agent does. Code Bubbles organized code into working sets and named recovery from interruptions as a future application~\cite{bubbles}. The evidence is mixed: in one document-management study, 3D interfaces were preferred but were not faster than 2D ones~\cite{cockburn}. This workshop has asked what extended reality could mean for future IDEs~\cite{xride} and has studied refactoring in virtual reality~\cite{vrrefactor}. Omo instead uses a familiar game world and a conventional IDE, and asks about returning to agent work rather than editing code in space.

Recent agent tools are closer still. ParallelPilot supports monitoring parallel AI coding sessions with logs, summaries, and an ambient dashboard~\cite{parallel}. Monet makes an agent's context visible and editable on a canvas~\cite{monet}. AgentCraft is also a Minecraft agent workspace: a lead agent plans work onto a wall and worker agents walk to desks~\cite{agentcraft}. Omo differs in what it puts in the world. AgentCraft embodies a team of agents; Omo places the user's own terminal sessions, running any agent or a plain shell, at locations the user builds. We do not claim to be the first world-based agent interface. Our question is different: whether a learned place helps a returning person reconnect with work that changed while they were gone.

\section{One Idea, Two Places}
\subsection{The Shared Design}
Both implementations separate the place from the process. A stable object in the scene, a computer block or a desk, stands for one named workspace on the user's machine. The workspace's terminals are owned by a tmux server on the host, not by the view, so closing the game or the browser does not stop the agents. The place adds a location to execution continuity. It does not summarize the work; it takes you to the real terminal, which may hold Hermes Agent, Claude Code, Codex, or a shell.

\subsection{Omo Space: The Idea at Full Scale}
\OS{} is an open-source Minecraft Java mod~\cite{omomc}. People build rooms, and blocks in those rooms are live terminals: a computer block holds a workspace, monitors show its terminal or a real web page, and a status lamp shows whether it is working, idle, or interrupted. Work keeps running when the player walks away or quits the game, and an interrupted command is never re-run automatically. Because building is the core of Minecraft, people can make a place that means something to them. Our own pipeline uses one room per stage (Fig.~\ref{fig:places}c), and about 30 members of our community report building their own worlds with it.

\subsection{Omo Theia: The Idea in an IDE}
\OT{} is a spatial IDE built on Eclipse Theia, released for macOS in November 2026~\cite{omoide}. It needs no game. Each project gets a themed room with desks, and each desk connects to terminals on the host (Fig.~\ref{fig:places}b). Rooms remember their layout and camera position, so a project looks the same every time. Its distinctive step is the short path from the place to the evidence: from a desk, the builder can open the project's Git patch (Fig.~\ref{fig:places}e) or the changed file in the editor. \OS{} shows how rich a place can be; \OT{} ties the place to what a reviewer must actually read.

\section{Lessons for IDE Builders}
Building the idea twice taught us four things that hold whether or not the spatial hypothesis turns out to be true.

\emph{Let sessions outlive views.} Agent terminals should be owned by the host, not the interface. Once they were, closing a window or the game stopped being a risk, and both front ends could share one backend.

\emph{Treat status as a prompt, not a verdict.} Our own ``working'' label outlived the agent's completion (Fig.~\ref{fig:places}d). A status label can tell you where to look, but only the output tells you what happened.

\emph{Link every place to evidence.} A familiar desk helps only if one step takes you to the patch and the changed files. In the worksheet example, it is the diff that reveals the missing export fix, not the room.

\emph{Never act on a guessed state.} Omo pastes a drafted instruction into the terminal without submitting it, and never re-runs interrupted commands, because the terminal might be showing a shell, a login, or an approval prompt instead of the agent.

\section{How We Will Test the Idea}
Our research question is: \emph{with the same access to the work, do persistent places help people resume AI-assisted tasks correctly after time away?} We plan a within-participant study comparing three interfaces that reach the same terminals, task text, changes, and editor: a list of well-named sessions, a stable 2D layout of workspaces, and an Omo room. The 2D condition matters because earlier work found 3D preferred but not faster~\cite{cockburn}. Participants, including novice and nonprofessional builders, would learn several small projects, leave, and return after a short break and again the next day to find agents that asked a question, made an incomplete fix, or changed something unexpected. A return counts as correct only if the person can state the original goal, explain what changed, and give an appropriate next instruction; unsafe instructions are recorded separately, so fast but wrong does not count. \OT{} suits this comparison because its review path can be kept identical across conditions.

\section{Discussion and Limitations}
A place has costs. It has to be built, kept stable, and navigated, and a familiar desk may make someone feel they understand work they have not checked. Minecraft makes places expressive but requires owning and running a game; Theia lowers that barrier but offers fewer ways to make a place personal. Moving work between desks or sharing a project may weaken the associations the idea relies on. The cognitive evidence also cuts both ways: walking through a doorway into a new room can cause forgetting, and in one experiment returning to the original room did not restore memory~\cite{radvansky}. Whether a place helps or hurts resumption is therefore an empirical question, not something the design can assume. Most importantly, we have no evidence yet that places help anyone but ourselves. A negative result would still be useful: it would mark the line between a compelling way to show agent work and a real aid to understanding it.

\section{Conclusion}
We built Omo because our agents kept working after we looked away, and coming back was the hard part. We now have the idea in two forms: a Minecraft world we work in every day, and an IDE that ties each place to the patch behind it. Building both showed us what a place gives, a quick way back to the right work, and what it does not, an account of what changed. Whether places help other people resume correctly, compared with well-named sessions or a stable 2D layout, is the question we want to answer next.

\balance
\section*{Availability}
\OS{} and \OT{} are open source~\cite{omomc,omoide}. All figure panels are captures from our own use of the two systems.

\bibliographystyle{IEEEtran}
\bibliography{references}
\end{document}